\documentclass[espf]{aa}
\usepackage{graphicx}

\titlerunning{}

\def\Mo{M_{\odot}}

\def\Ro{R_{\odot}}

\begin{document}

\title{An investigation of the close environment of $\beta$ Cep with the VEGA/CHARA interferometer}

\titlerunning{An investigation of the close environment of $\beta$ Cep with the VEGA/CHARA interferometer}
\authorrunning{Nardetto et al. }

\author{N. Nardetto \inst{1}, D. Mourard \inst{1}, I. Tallon-Bosc\inst{2}, M. Tallon\inst{2}, P. Berio \inst{1}, E. Chapellier \inst{1}, D. Bonneau \inst{1}, O.~Chesneau \inst{1}, P. Mathias \inst{3}, K. Perraut \inst{4}, P. Stee \inst{1}, A. Blazit \inst{1}, J. M. Clausse \inst{1}, O. Delaa \inst{1}, A. Marcotto \inst{1}, F.~Millour \inst{5}, A. Roussel \inst{1}, A. Spang\inst{1}, H. McAlister\inst{6,7}, T. ten~Brummelaar\inst{7}, J. Sturmann\inst{7}, L. Sturmann\inst{7}, N. Turner\inst{7}, C. Farrington\inst{7} and P.J. Goldfinger\inst{7}}

\institute{Laboratoire Fizeau, UNS/OCA/CNRS UMR6525, Parc Valrose, 06108 Nice Cedex 2, France \and Universit\'e de Lyon, 69003 Lyon, France; Universit\'e Lyon 1, Observatoire de Lyon, 9 avenue Charles Andr\'e, 69230 Saint Genis Laval, France; CNRS/UMR 5574, Centre de Recherche Astroph. de Lyon; Ecole Normale Sup\'erieure,
69007 Lyon, France;  \and Laboratoire d'Astrophysique de Toulouse-Tarbes, Universit\'e de 
Toulouse, CNRS, 57 Avenue dÕAzereix, F-65000 Tarbes \and Laboratoire d'Astrophysique de Grenoble (LAOG), Universit\'e Joseph-Fourier, UMR 5571 CNRS, BP 53,
38041 Grenoble Cedex 09, France \and  Max-Planck-Institut f\"ur Radioastronomie, Auf dem H\"ugel 69, 53121 Bonn, Germany
\and Georgia State University, P.O. Box 3969, Atlanta GA 30302-3969, USA 
\and CHARA Array, Mount Wilson Observatory, 91023 Mount Wilson CA, USA}


\date{Received ... ; accepted ...}

\abstract{High-precision interferometric measurements of pulsating stars help to characterize their close environment. In 1974, a close companion was discovered around the pulsating star $\beta$ Cep using the speckle interferometry technique and features at the limit of resolution ($20$ milli-arcsecond or mas) of the instrument were mentioned that may be due to circumstellar material. $\beta$ Cep has a magnetic field that might be responsible for a spherical shell or ring-like structure around the star as described by the MHD models.} {Using the visible recombiner VEGA installed on the CHARA long-baseline interferometer at Mt. Wilson, we aim to determine the angular diameter of $\beta$ Cep and resolve its close environment with a spatial resolution up to 1 mas level.} {Medium spectral resolution (R=6000) observations of $\beta$ Cep were secured with the VEGA instrument over the years 2008 and 2009. These observations were performed with the S1S2 (30m) and W1W2 (100m) baselines of the array.} {We investigated several models to reproduce our observations. A large-scale structure of a few mas is clearly detected around the star with a typical flux relative contribution of $0.23 \pm 0.02$. Our best model is a co-rotational geometrical thin ring around the star as predicted by magnetically-confined wind shock models. The ring inner diameter is $8.2 \pm 0.8$ mas and the width is $0.6 \pm 0.7$~mas. The orientation of the rotation axis on the plane of the sky is PA$=60 \pm 1$ deg, while the best fit of the mean angular diameter of $\beta$ Cep gives $\Phi_{\mathrm{UD}}[V]=0.22 \pm 0.05$~mas.  Our data are compatible with the predicted position of the close companion of $\beta$ Cep.} {These results bring additional constraints on the fundamental parameters and on the future MHD and asteroseismological models of the star.}


\maketitle

\section{Introduction}\label{s_Introduction}

The prototype of the $\beta$ Cephei class of pulsating stars, $\beta$~Cep (HD205021, spectral type B2III, V=3.2), is a massive ($M=12\pm 1 \Mo$, Donati et al. 2001) pulsating star with a period of approximatively $P=0.1905$ days (Kukarkin et al. 1971). The radial-velocity variations of this star were first detected by Frost (1902). Five frequencies have been derived from a long-term spectroscopic study of $\beta$ Cep, two of which were identified as a radial and non radial mode (Telting et al. 1997). The radius of $\beta$ Cep is $R=6.5 \pm1.2 \Ro$ (Donati et al. 2001), and its distance is $d = 210 \pm 13 pc$ (Van Leeuwen 2007). This corresponds to a limb darkened angular diameter of $0.29\pm0.06$ mas. 

The source $\beta$ Cep is actually known as a tertiary system ("Washington Double Star Catalog"\footnote{http://ad.usno.navy.mil/wds/}). The pulsating star is a member of a visual pair (STF 2866 AB, $a=13.4$", $\Delta m$ = 5.4). The primary is a close pair (LAB6 Aa,Ab, $\Delta m$ = 3.4), first resolved by speckle interferometry at a separation of approximatively 0.25" (Labeyrie 1970, Gezari et al. 1972). Since 1971.48, this system has been extensively observed by speckle interferometry (see  the "Fourth Catalog of Interferometric Measurements of Binary Stars"\footnote{http://ad.usno.navy.mil/wds/int4.html}) and has led to the computation of the current orbit with a period of $83$ years  (Andrade 2006) found in the "Sixth Catalog of Orbits of Visual Binary Stars"\footnote{http://ad.usno.navy.mil/wds/orb6.html}. In 1974, Labeyrie et al. noted anisotropic features in the power spectrum of the speckled images, which might indicate a large-scale structure at the limit of resolution (20 mas) of the Palomar $5$m telescope. Recently, the technique of spectroastrometry was used to disentangle the component spectra of the speckle binary (Schnerr et al. 2006, Wheelwright 2009). These authors conclude that the H$\alpha$ emission is certainly due to the close companion of $\beta$ Cep, which appears to be a classical Be star.

A magnetic field strength of $360$G was measured by Henrichs et al. (2000). The magnetic axis is almost 90 degrees from the rotational axis, whose inclination is 60 degrees compared to the line-of-sight (Abt et al. 2002). The rotation period of the star is about 12 days (Donati et al. 2001). If the magnetic field is strong enough the wind can be fully magnetically confined in a region close to the star (Babel \& Montmerle 1997). In this case, the stellar wind particles originating in both magnetic poles and forced to follow the magnetic field lines collide at the magnetic equator, creating a decretion ring or circumstellar clouds. The magnetic field could confine the wind of $\beta$ Cep from 2$R_{\star}$ up to a distance of about 8 to 9 $R_{\star}$ (Donati et al. 2001). Of course, the apparent geometry of the ring will then be modulated by the rotation phase of the star. Later work based on dynamical modeling of the wind-magnetic-field interaction, however, indicates that the thick ring (in the X-ray domain) predicted by Babel \& Montmerle (1997) would be unlikely to form around a star such as $\beta$ Cep  (Gagn\'e et al. 2005; ud-Doula \& Owocki 2002; Townsend \& Owocki 2005). This hypothesis has recently been observationally supported  by Favata et al. (2009), who did not find any strong rotational modulation  in the X-ray emission as predicted by Donati et al. (2001), which seems to point toward a spherical shell of X-ray emitting plasma around the star located between 5$R_{\star}$  to 7$R_{\star}$ (for He-like triplets). In this context, long-baseline interferometry is in a unique position to derive the angular diameter of the star and the mas scale geometric structure of its close environment.

The Center for High Angular Resolution Astronomy (CHARA) of Georgia State University operates an optical interferometric array located at Mount Wilson Observatory (ten Brummelaar et al. 2005). It is formed by six telescopes placed in pairs on the arms of a Y-shaped configuration. It yields 15 baselines ranging from $34$ to $331$m. The VEGA instrument works in the visible domain and provides a maximum angular resolution of $0.3$ mas (Mourard et al. 2009). In Sect. 2, we present the VEGA observations of $\beta$ Cep and the data reduction process. Then, in Sect. 3, we propose several models of $\beta$ Cep composed of the star itself and additional geometrical structures such as a Gaussian, a companion, or a ring. The last section is devoted to a discussion and a few conclusions.   

\section{Observations and data processing}

\begin{figure}[htbp]
\begin{center}
\resizebox{\hsize}{!}{\includegraphics[clip=true]{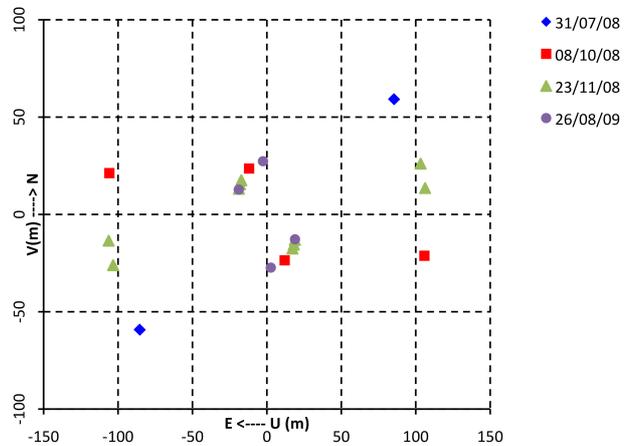}}
\end{center}
\caption{ (u-v) plane coverage of VEGA observations} \label{Figuv}
\end{figure}

\begin{table*}
\begin{center}
\caption[]{Summary of observations and measured visibilities.
\label{Tab_log}}
\begin{tabular}{ccccccccccccc}
\hline \hline \noalign{\smallskip}
date           &     JD$_c^\mathrm{a}$                &  AH        &   $\lambda$      &  $\Delta_\lambda$            &      $B_{\mathrm{p}}$        & Arg.           & S/N &  $V^2 \pm V^2_\mathrm{stat} \pm  V^2_\mathrm{syst}$  &   CAL$^\mathrm{b}$ &    $\phi_{\mathrm{rot}}^\mathrm{c}$ & e$_{\mathrm{r}}^\mathrm{d}$ \\
{\it dd/mm/yy}                   &   [days]            & [hour]    &               [nm]                                    &       [nm]                      &     [m]     &         [deg]      &                                                                                             &                       &             &                &                     \\
                   \hline
31/07/08	&	4677.5	&	3.17	&	635.00	&	20.00		&	104	&	55	&	15	$\pm$	6	&	0.684	$\pm$	0.061	$\pm$	0.020	&	C1-S	&	0.62	 & 1.37\\
31/07/08	&	4677.5	&	3.18	&	655.00	&	20.00		&	104	&	55	&	35	$\pm$	8	&	0.611	$\pm$	0.083	$\pm$	0.016	&	C1-S	&	0.62	& 1.37\\
\hline
08/10/08	&	4746.5	&	-0.03	&	528.75	&	12.50		&	108	&	101	&	47	$\pm$	10	&	0.671	$\pm$	0.064	$\pm$	0.030	&	C1-S-C1	&	0.37	& 1.46\\
08/10/08	&	4746.5	&	1.40	&	528.75	&	12.50		&	26	&	-27	&	93	$\pm$	33	&	0.563	$\pm$	0.048	$\pm$	0.001	&	C1-S	&	0.37	& 1.46\\
08/10/08	&	4746.5	&	-0.03	&	541.50	&	13.00		&	108	&	101	&	41	$\pm$	14	&	0.578	$\pm$	0.049	$\pm$	0.025	&	C1-S-C1	&	0.37	& 1.46\\
08/10/08	&	4746.5	&	1.40	&	541.50	&	13.00		&	26	&	-27	&	86	$\pm$	28	&	0.658	$\pm$	0.050	$\pm$	0.001	&	C1-S	&	0.37& 1.46	\\
08/10/08	&	4746.5	&	-0.03	&	680.00	&	20.00		&	108	&	101	&	57	$\pm$	10	&	0.738	$\pm$	0.065	$\pm$	0.020	&	C1-S-C1	&	0.37	& 1.46\\
08/10/08	&	4746.5	&	1.40	&	680.00	&	20.00		&	26	&	-27	&	72	$\pm$	26	&	0.586	$\pm$	0.028	$\pm$	0.001	&	C1-S-C1	&	0.37	& 1.46\\
08/10/08	&	4746.5	&	-0.03	&	700.00	&	20.00		&	108	&	101	&	79	$\pm$	29	&	0.674	$\pm$	0.051	$\pm$	0.017	&	C1-S-C1	&	0.37	& 1.46\\
08/10/08	&	4746.5	&	1.40	&	700.00	&	20.00		&	26	&	-27	&	101	$\pm$	19	&	0.597	$\pm$	0.029	$\pm$	0.001	&	C1-S-C1	&	0.37	& 1.46\\
\hline
23/11/08	&	4792.5	&	3.06	&	507.50	&	15.00		&	25	&	-44	&	51	$\pm$	9	&	0.545	$\pm$	0.049	$\pm$	0.001	&	C2-S	&	0.21	& 4.02\\
23/11/08	&	4792.5	&	3.52	&	507.50	&	15.00		&	24	&	-49	&	15	$\pm$	5	&	0.604	$\pm$	0.107	$\pm$	0.002	&	C2-S	&	0.21	& 4.02\\
23/11/08	&	4792.5	&	4.04	&	507.50	&	15.00		&	23	&	-55	&	28	$\pm$	6	&	0.563	$\pm$	0.050	$\pm$	0.001	&	C2-S	&	0.21& 4.02	\\
23/11/08	&	4792.5	&	1.28	&	660.00	&	20.00		&	107	&	83	&	41	$\pm$	10	&	0.627	$\pm$	0.152	$\pm$	0.017	&	C1-S	&	0.21	& 4.02\\
23/11/08	&	4792.5	&	1.77	&	660.00	&	20.00		&	107	&	76	&	58	$\pm$	6	&	0.489	$\pm$	0.053	$\pm$	0.012	&	C2-S	&	0.21	& 4.02\\
23/11/08	&	4792.5	&	3.06	&	660.00	&	20.00		&	25	&	-44	&	108	$\pm$	35	&	0.590	$\pm$	0.062	$\pm$	0.001	&	C2-S	&	0.21	& 4.02\\
23/11/08	&	4792.5	&	3.52	&	660.00	&	20.00		&	24	&	-49	&	34	$\pm$	11	&	0.639	$\pm$	0.088	$\pm$	0.001	&	C2-S	&	0.21	& 4.02\\
23/11/08	&	4792.5	&	4.04	&	660.00	&	20.00		&	23	&	-55	&	47	$\pm$	14	&	0.585	$\pm$	0.064	$\pm$	0.001	&	C2-S	&	0.21	& 4.02\\
23/11/08	&	4792.5	&	1.28	&	680.00	&	20.00		&	107	&	83	&	66	$\pm$	17	&	0.583	$\pm$	0.055	$\pm$	0.015	&	C1-S	&	0.21	& 4.02\\
23/11/08	&	4792.5	&	1.77	&	680.00	&	20.00		&	107	&	76	&	66	$\pm$	18	&	0.614	$\pm$	0.075	$\pm$	0.014	&	C2-S	&	0.21	& 4.02\\
23/11/08	&	4792.5	&	3.06	&	680.00	&	20.00		&	25	&	-44	&	134	$\pm$	31	&	0.686	$\pm$	0.081	$\pm$	0.001	&	C2-S	&	0.21	& 4.02\\
23/11/08	&	4792.5	&	3.52	&	680.00	&	20.00		&	24	&	-49	&	38	$\pm$	15	&	0.702	$\pm$	0.057	$\pm$	0.001	&	C2-S	&	0.21	& 4.02\\
23/11/08	&	4792.5	&	4.04	&	680.00	&	20.00		&	23	&	-55	&	72	$\pm$	9	&	0.600	$\pm$	0.066	$\pm$	0.001	&	C2-S	&	0.21	& 4.02\\
\hline
26/08/09	&	5068.5	&	-0.63	&	491.25	&	12.50		&	27	&	-6	&	47	$\pm$	15	&	0.435	$\pm$	0.033	$\pm$	0.001	&	C2-S-C2	&	0.20	& 3.24\\
26/08/09	&	5068.5	&	-0.63	&	503.75	&	12.50		&	27	&	-6	&	51	$\pm$	9	&	0.468	$\pm$	0.023	$\pm$	0.001	&	C2-S-C2	&	0.20	& 3.24\\
26/08/09	&	5068.5	&	-0.63	&	660.00	&	20.00		&	27	&	-6	&	39	$\pm$	10	&	0.556	$\pm$	0.040	$\pm$	0.001	&	C2-S-C2	&	0.20	& 3.24\\
26/08/09	&	5068.5	&	4.11	&	660.00	&	20.00		&	23	&	-56	&	46	$\pm$	8	&	0.655	$\pm$	0.036	$\pm$	0.001	&	C2-S-C2	&	0.20	& 3.24\\
26/08/09	&	5068.5	&	-0.64	&	680.00	&	20.00		&	27	&	-5	&	59	$\pm$	15	&	0.574	$\pm$	0.019	$\pm$	0.001	&	C2-S-C2	&	0.20	& 3.24\\
26/08/09	&	5068.5	&	4.12	&	680.00	&	20.00		&	23	&	-56	&	70	$\pm$	18	&	0.679	$\pm$	0.034	$\pm$	0.001	&	C2-S-C2	&	0.20	& 3.24\\
\hline \noalign{\smallskip}
\end{tabular}
\end{center}
\begin{list}{}{}
\item[$^{\mathrm{a}}$] Julian date of observation defined by $JD_{\mathrm{c}}=JD-2450000$  
\item[$^{\mathrm{b}}$] Observation sequence. C1, C2 and S are for calibrator 1 (HD192907), calibrator 2 (HD214734) and science star respectively. 
\item[$^{\mathrm{c}}$] The reference Julian date ($T_0=2451238.15$ days) and the rotation
period ($P=12.00092$ days) used to compute the phase are from Donati et al. (2001).  The phase 0 is the phase at which the longitudinal magnetic
field is at a maximum.
\item[$^{\mathrm{d}}$] The elongation of the ring derived from Figure 2 of Favata et al. (2009). See also Fig. \ref{Fig_elongation}.
\end{list}
\end{table*}

The VEGA instrument works in the visible band [$0.45\mu$m; $0.85\mu$m] and benefits from three spectral resolutions. The medium (6000) and high (30 000) spectral resolutions are well-suited to kinematic analysis of the interferometric signal, while low (1700) and medium resolutions are well-suited to absolute visibility studies and are also well adapted to the study of binaries or multiple systems. We used the medium spectral resolution to observe $\beta$ Cep. 

We obtained observations on different dates (31/07/08, 08/10/08, 23/11/08, and 26/08/09) and different baselines (S1S2, W1W2) to ensure a good (u-v) plane coverage (Fig.~\ref{Figuv}). Some of these observations were done remotely from Grasse observatory (Clausse 2008). VEGA is equipped with two photon-counting detectors looking at two different spectral bands (typically $0.55\mu$m  and $0.70\mu$m, respectively). In medium resolution, the two spectral channels have a typical wavelength range of $30$ and $45$nm. We subdivided these spectral domains in $2$ parts in order to calculate the raw (or instrumental) visibilities of the science and calibrator stars, by measuring the ratio of the high to low-frequency energy of the averaged Fourier spectrum of the fringes visibility (Mourard et al. 2009).

An absolute calibration of the science visibilities was performed using references stars HD 192907 and HD 214734 selected with the {\it SearchCal}\footnote{Available at http://www.jmmc.fr/searchcal} software provided by the Jean-Marie Mariotti Center (JMMC), which gives computed photometric angular diameters $\Phi_{\mathrm{LD}}=0.357\pm0.025$ mas for HD 192907 and $\Phi_{\mathrm{LD}}=0.340\pm0.023$~mas for HD 214734 (Bonneau et al. 2006). 

The log and sequence of observations are presented in Table \ref{Tab_log}.  For each calibrated visibility, the statistical and systematic calibration errors are given separately. In most cases, the systematic calibration error is negligible compared to the statistical one. In the model fitting we only consider the statistical uncertainties.

\section{Model fitting and strategy}

To perform the model fitting we used a tool, {\it LITpro}, which is currently being developed in the framework of the JMMC\footnote{http://www.jmmc.fr/litpro} (Tallon-Bosc et al. 2008). The fitting engine is based on a modified Levenberg-Marquardt algorithm combined with the trust regions method. The software provides a user-expandable set of geometrical elementary models of the object, able to be combined as building blocks. 

All our observations (whatever the baseline, the wavelength, the time, and date of observations) give visibilities ranging from $0.435$ to $0.738$. It seems to indicate that the instrument is resolving a large-scale feature in the close environment of $\beta$ Cep. Using the model building blocks provided by {\it LITpro} we construct a geometrical view of the target step by step. We focus in particular on the relative flux contributions of the star and its close environment, quantities that are indeed strongly constrained by VEGA interferometric observations. Models and fitting results are summarized in Table \ref{Tab_sum}: (1) star + Gaussian, (2) star + circular ring, (3) star + Gaussian  + companion at 170 mas (the relative flux contribution of the different geometric components are indicated, respectively) and, (4) star + peculiar ring. The symbol $\star$ indicates that the parameter is fixed. 

\subsection{A circumstellar envelope}

\begin{figure}[htbp]
\begin{center}
\resizebox{\hsize}{!}{\includegraphics[clip=true]{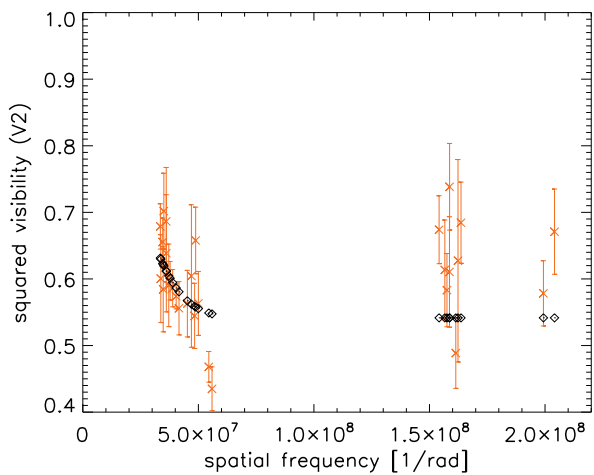}}
\end{center}
\caption{Visibility of a model  composed of $\beta$ Cep (unresolved) and a gaussian circumstellar envelope (diamonds) together with VEGA observations (crosses). All data are reported in one spatial frequency dimension. } \label{Fig_PS_GAUSS}
\end{figure}

We consider a model composed of an uniform disk and a Gaussian circumstellar envelope. The best-fit results in an unresolved central object. The envelope flux contributions in the optical range of VEGA is $0.26\pm0.02$. The full-width at half-maximum (hereafter fwhm) of the Gaussian is $4.0 \pm 0.4$ mas and the reduced $\chi^2$ is 2.5 (Fig. \ref{Fig_PS_GAUSS}). The fit is better than for a single-star model ($\chi^2$ of 37), which indicates that there is indeed a large-scale structure around $\beta$ Cep. We now try to remove the residual discrepancies (especially for the W1W2 baseline), by including the close companion of $\beta$ Cep in the model.  

Following Favata et al. (2009) we tested the hypothesis of a spherical shell around the star by simply considering a circular ring. Our data are consistent with this kind of model with a reduced $\chi^2$ of $1.4$.  We find the following parameters.  The star is again unresolved and  the relative flux contribution of the ring is $0.22\pm0.03$; however, the geometry of the ring is not really constrained with an inner angular diameter of $2.7\pm1.9$ mas and a width of $1.3\pm0.1$~mas. 

\subsection{A close companion at about 170 mas}

Using the orbital parameters of the close companion of $\beta$ Cep presented in Andrade (2006), we added the close unresolved companion of  $\beta$ Cep in our model of the Gaussian circumstellar envelope in order to evaluate its impact on the fitting process. Averaging the four dates of observations, we find $x_c~=~122.8$~mas and $y_c~=~123.2$~mas, where $x_c$ and $y_c$ are the coordinates of the close companion of $\beta$ Cep. Here, $x$ (resp.~$y$) is counted positive toward the east (resp. north). A recent determination by Mason et al. (2009) gives $\Delta m=3.4$, which corresponds to a flux contribution to the total flux of $4\%$.  We found from the best fit that $\beta$ Cep is still unresolved and an fwhm for the Gaussian of $3.9\pm0.4$mas, consistent with what was obtained in the previous section.  The flux contribution of the Gaussian and the binary (compared to the total flux) are $0.26 \pm 0.03$ and $0.01\pm0.01$, respectively. The reduced $\chi^2$ is $2.5$. The presence of the companion induces a sinusoidal modulation of amplitude about $0.03$ in the visibility curve (in the $v$ direction), but it does not help reproduce observations satisfactorily. Our data are thus not really sensitive to the position of the companion. Therefore, the close companion is neglected in the remainder of the analysis.

\subsection{A peculiar ring around $\beta$ Cep}

As mentioned in Sect. 1, the hypothesis of a ring surrounding $\beta$ Cep is supported by MHD codes. In a first approximation we can assume that the ring is rotating about an axis that is in the plane of the sky (Fig. 2 of Favata et al. 2009). Consequently, the ring is modulated by the rotation of the star in such a way that it is alternatively in a pole-on and edge-on geometry. The parameters of the model we use are defined in the {\it LITpro} software as 

 \begin{figure}[htbp]
\begin{center}
\resizebox{\hsize}{!}{\includegraphics[clip=true]{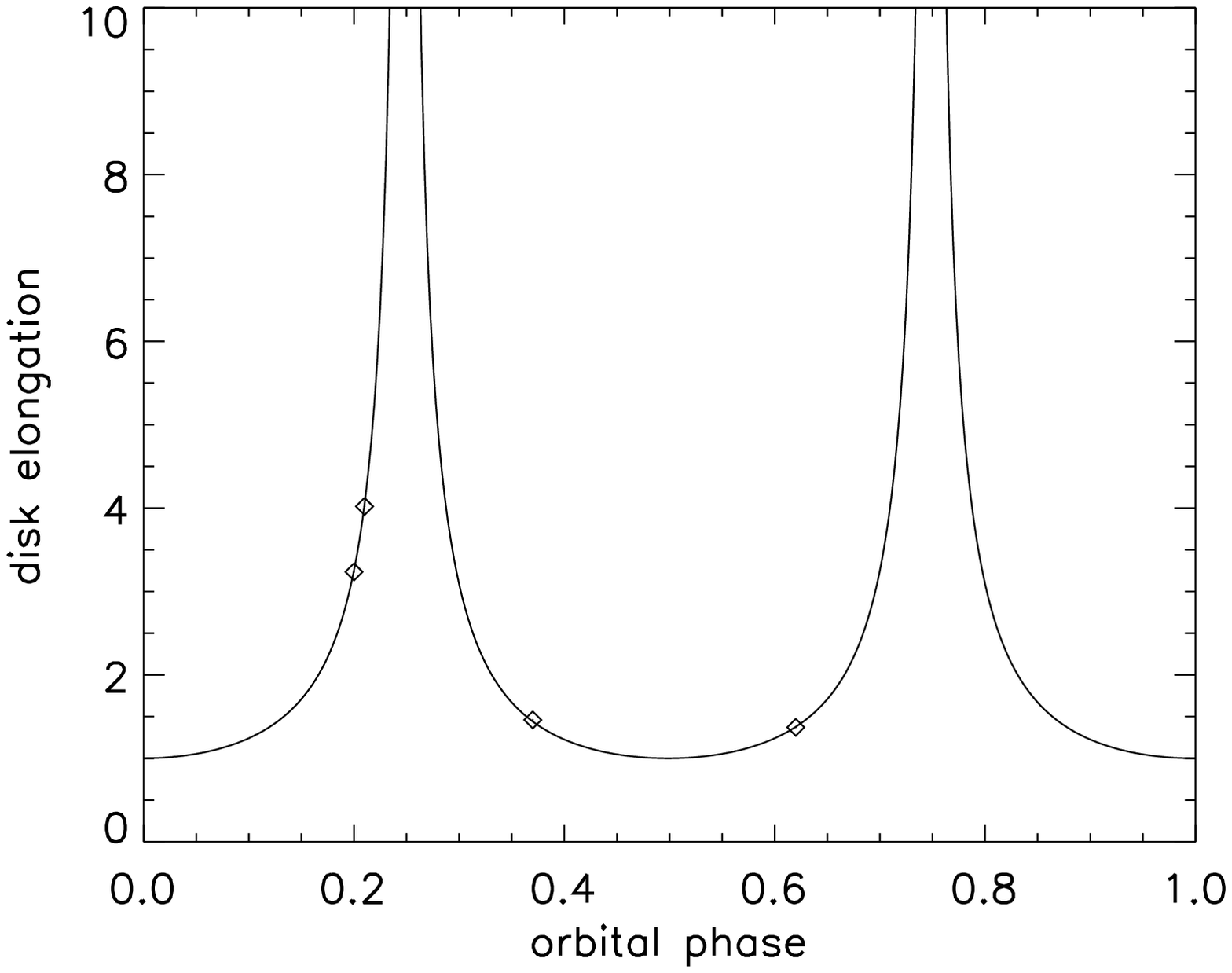}}
\end{center}
\caption{Elongation of the ring $e_{\mathrm{r}}$ as a function of the orbital phase. Diamond represents the predicted elongation at the observational dates of VEGA (see Tab. \ref{Tab_log}). $e_{\mathrm{r}}=1$ and $e_{\mathrm{r}}=\infty$ correspond to a pole-on and edge-on geometry, respectively.} \label{Fig_elongation}
\end{figure}

\begin{itemize}
\item $\Phi_\star$, the angular diameter of $\beta$ Cep [in mas]
\item $f_\mathrm{r}$, the relative flux contribution of the ring in the optical range of VEGA, compared to the total flux 
\item $\Phi_\mathrm{r}$, the ring inner diameter [in mas]
\item $w_\mathrm{r}$, the width of the ring [in mas]
\item the elongation defined as $e_{\mathrm{r}}=\frac{\Phi_\mathrm{r}}{\Phi_\mathrm{rmin}}=\frac{\omega_\mathrm{r}}{\omega_\mathrm{rmin}}$ where $\Phi_{\mathrm{rmin}}$ and $\omega_{\mathrm{rmin}}$ are the apparent angular inner radius and width of the ring along the minor axis. 
\item  PA, the orientation of the major axis of the ring on the plane of the sky. It is measured in degrees, from the positive vertical semi-axis (i.e. north direction) towards to the positive horizontal semi-axis (i.e. east direction).    
\end{itemize}

 \begin{figure*}[htbp]
\begin{center}
\resizebox{0.48\hsize}{!}{\includegraphics[clip=true]{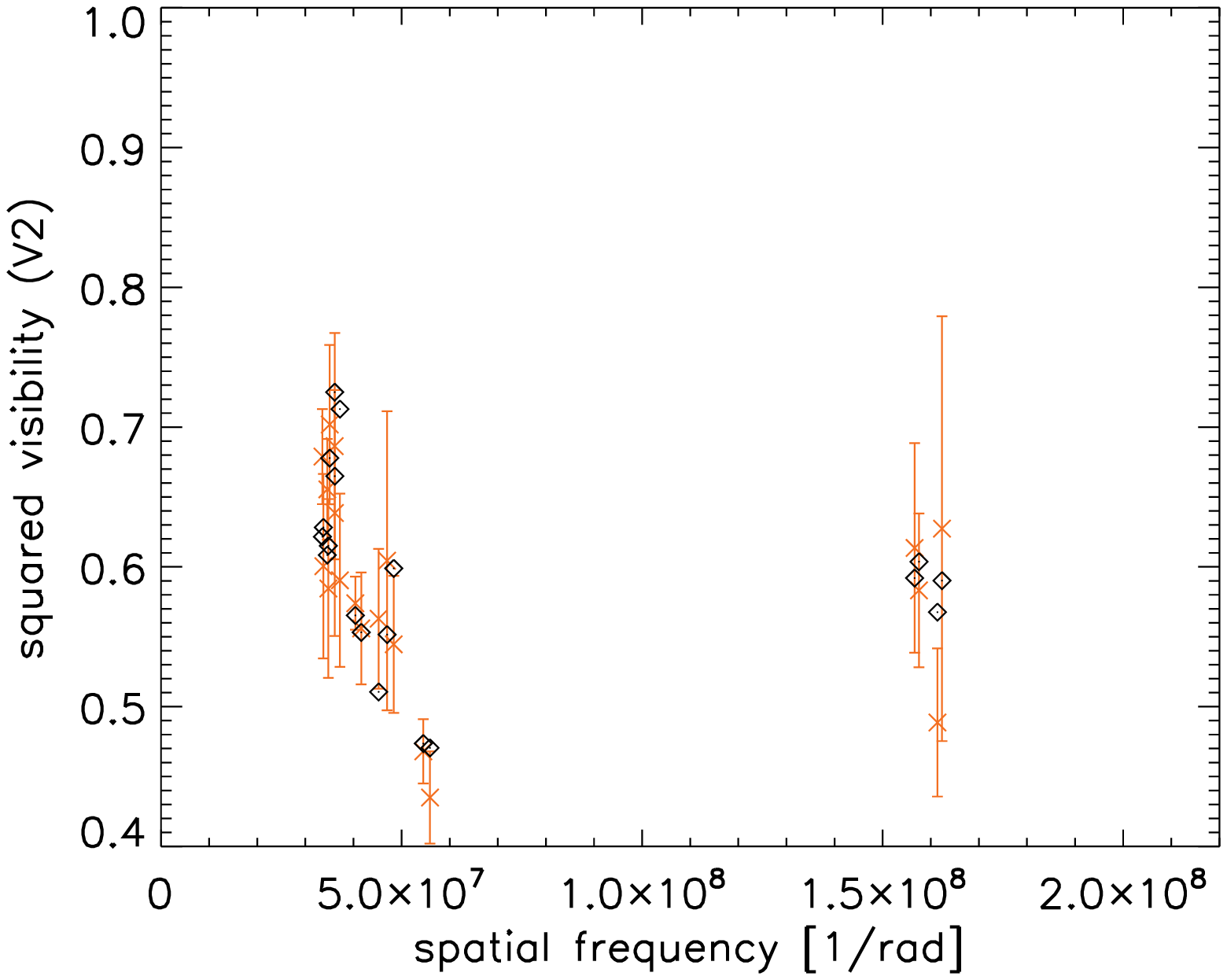}}
\resizebox{0.48\hsize}{!}{\includegraphics[clip=true]{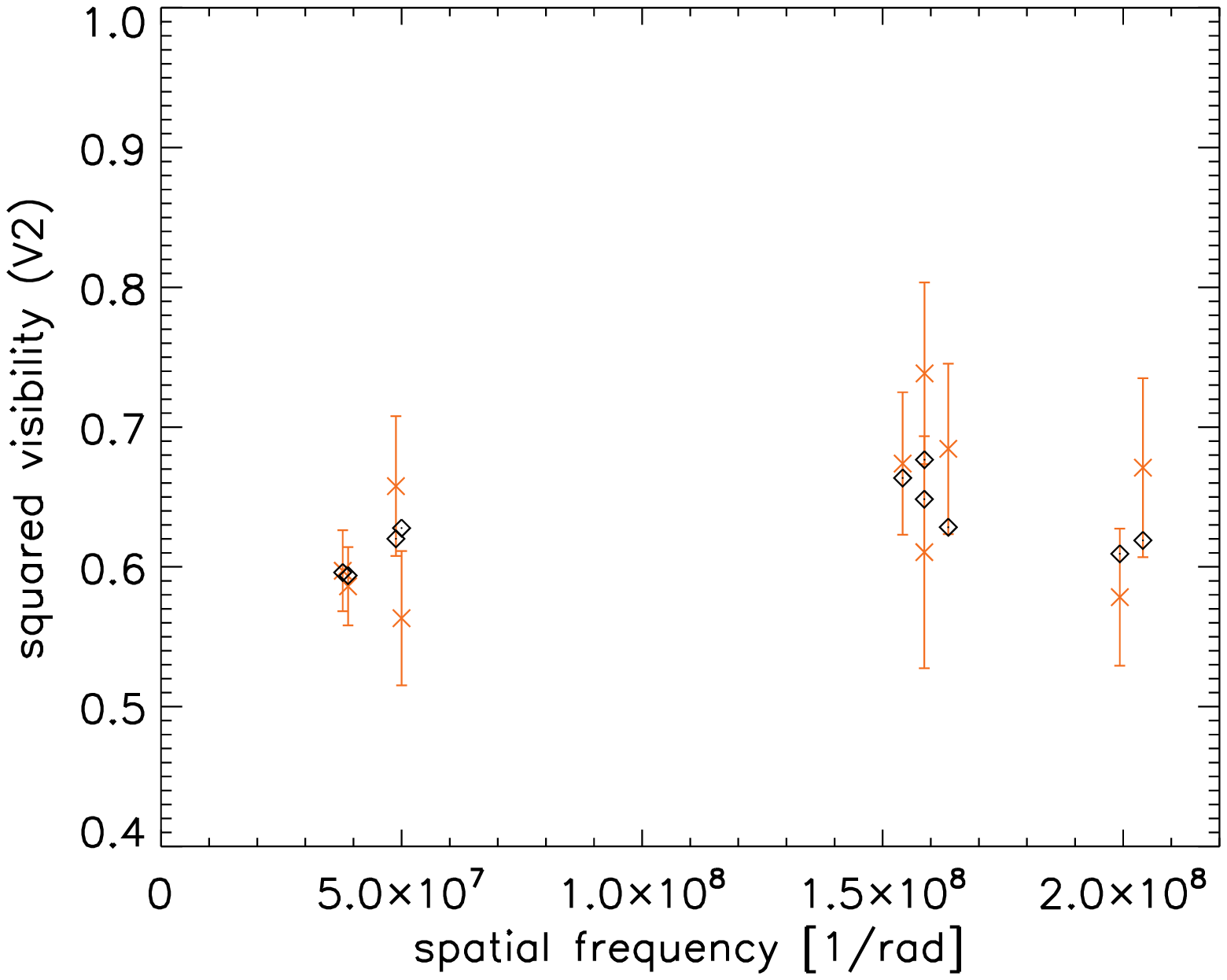}}
\end{center}
\begin{center}
\resizebox{0.40\hsize}{!}{\includegraphics[clip=true]{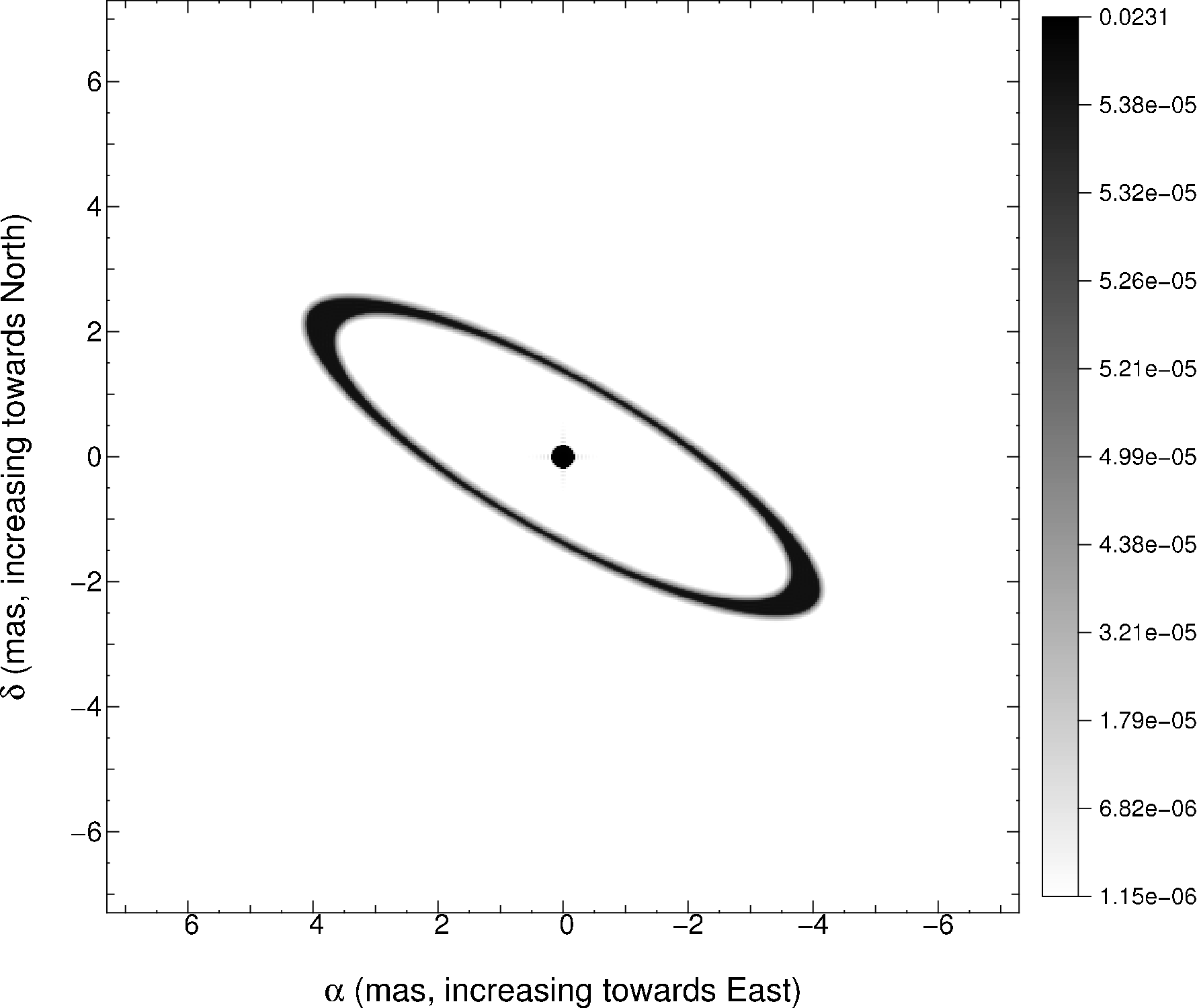}}
\hspace{2cm}
\resizebox{0.40\hsize}{!}{\includegraphics[clip=true]{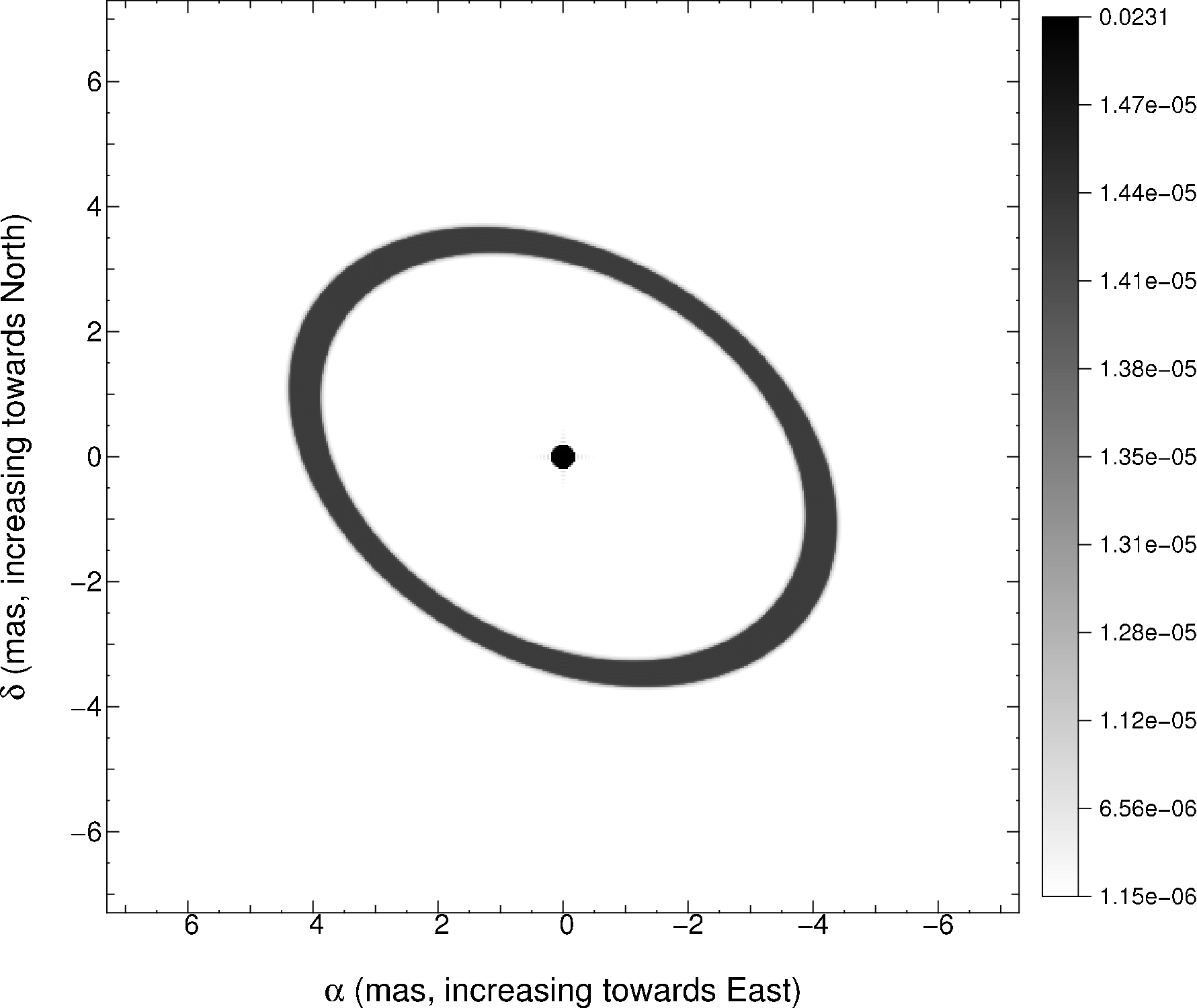}}
\end{center}
\caption{{\bf Left}: Same as Fig. \ref{Fig_PS_GAUSS} but with a model composed of $\beta$ Cep and an edge-on ring. Most data are from S1S2 short baseline (see nights 23/11/08 and 26/08/09), which help constrain the geometry of the ring and, in particular, its orientation on the sky. The image of the model is also provided. {\bf Right}: The same but with a model composed of $\beta$ Cep and a pole-on ring. In this case, we have more data for the W1W2 long baseline (see nights 31/07/08 and 08/10/08), which helps constrain the angular diameter of the $\beta$ Cep. } \label{Fig_PS_RING}
\end{figure*}

From the ephemeris from Donati et al. (2001), we calculated the rotation phases corresponding to our date of observation (Table \ref{Tab_log}). Then, using Figure 2 in Favata et al. (2009), we derived the elongation $e_{\mathrm{r}}$ as a function of the rotation phase (Fig. \ref{Fig_elongation}).  For the four nights of observation, we obtain the following elongations indicated in Table \ref{Tab_log}. We basically have two sets of data with an almost similar elongation of the ring. The data corresponding to the nights  23/11/08 and 26/08/09 were analyzed together in the fitting process considering a mean elongation of $e_{\mathrm{r}}=3.63$ (ring almost edge-on). Similarly, for nights 31/07/08 and  08/10/08, we consider a mean elongation of $e_{\mathrm{r}}=1.42$ (ring almost pole-on). The elongation parameter is fixed in the following, because the data sample is not large enough to constrain all parameters at the same time.  We now describe the strategy we used to optimize the fitting process. The parameters $\Phi_\star$, $\Phi_\mathrm{r}$, $w_\mathrm{r}$, and PA are supposed to be the same in the edge-on and pole-on geometries, and they are linked during the fitting process (an interesting possibility provided by the {\it LITpro} software). The relative flux $f_{\mathrm{r}}$ is fitted in both cases, because the brightness of the ring might change with the elongation. We also neglect the radius variation of the star with the pulsation phase (about $1$\%), which is below the capabilities of the VEGA instrument. The reduced $\chi^2$ is $0.9$. Fixed parameters and results are presented in Table \ref{Tab_sum}, while the fitting quality can be appreciated in Fig. \ref{Fig_PS_RING}.

\begin{figure}[htbp]
\begin{center}
\resizebox{\hsize}{!}{\includegraphics[clip=true]{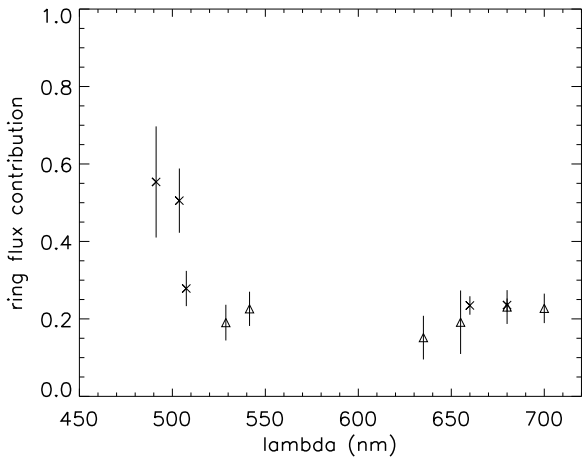}}
\end{center}
\caption{Ring flux contribution as a function of the wavelength. Crosses and triangles correspond to the edge-on and pole-on ring geometry respectively.} \label{Fig_ratio}
\end{figure}

An implicit assumption is that the ring flux contribution is independent of wavelength. For each visibility measurement $V(\lambda)$ of Table 1, the ring flux contribution $f_\mathrm{r}(\lambda)$ is defined as $\frac{V(\lambda)-V_\star}{V_\mathrm{r}-V_\star}$, where $V_\star$ is the modeled visibility of the central star, and $V_\mathrm{r}$ of the ring.  The values of $f_\mathrm{r}$ with the same wavelength are averaged. Results are presented in Fig. \ref{Fig_ratio}. The optical emission in the ring-like gaseous structure is mainly due to Rayleigh diffusion at short wavelength ($f_\mathrm{r} (\lambda)$ is decreasing), while the Thomson scattering might be dominant for wavelengths longer than $\lambda > 510$nm ($f_\mathrm{r} (\lambda)$ is constant).

\begin{table*}
\begin{center}
\caption[]{Models presented in this paper. }
\label{Tab_sum}
\begin{tabular}{|c|c|c|c|cccc|cc|c|}
\hline \hline \noalign{\smallskip}
  Model                                      &    $\Phi_\star$      & flux contributions                                                                                            &  fwhm                 & $\Phi_\mathrm{r}$ &  $w_\mathrm{r}$ & $e_{\mathrm{r}}$ & PA & $x_c$ & $y_c$ & $\chi_{\mathrm{red}}^2$   \\ 
                                                    &   [mas]      &                                                                                              &  [mas]                 & [mas] &   [mas] &  & [deg] &  [mas] &  [mas] &    \\ 
\hline

1      &  0                          &                               $0.26_{\pm0.02}$                                         &    $4.0_{\pm 0.4}$ &   -                            &              -               &     -       &    -     &    -     &     -       &  $2.5$ \\
2      &  0                          &                               $0.22_{\pm0.03}$                                       &           -                           &  $2.7_{\pm1.9}$    &   $1.3_{\pm1.1}$  &     $1.00^{\star}$       &    -     &    -    &     -     &  $1.4 $ \\
3      &  0                          &                               $0.26_{\pm0.03}$, $0.01_{\pm 0.01}$        &    $3.9_{\pm 0.4}$ &   -                            &              -               &     -       &    -     &    $122.8^{\star}$     &     $123.2^{\star}$     &  $2.5$ \\

\hline
        &     &                               $0.22_{\pm0.01}$                                       &      -  &       &    &     $3.63^{\star}$       &         &    -    &     -     &  \\
4      &  $0.22_{\pm0.05}$ &                                                                     &     -    &  $8.2_{\pm0.8}$  &   $0.6_{\pm0.7}$    &            &   $60_{\pm1}$    &  -      &  -      &   $0.9$\\
        &     &                               $0.17_{\pm0.01}$                                       &      -  &        &      &     $1.42^{\star}$       &        &    -    &     -     &   \\
\hline \noalign{\smallskip}
\end{tabular}
\end{center}
\end{table*}

\section{Discussion and conclusion}

We progressively improved our models by exploring several hypothesis. Models with a large-scale structure of several mas around $\beta$ Cep have a reduced $\chi^2$ between 15 to 38 times lower than the uniform disk hypothesis. The mean relative flux contribution of this large-scale structure over all the models presented in this paper is $0.23 \pm 0.02$. This is certainly our most important result. 

Our best model (reduced $\chi^2$ of $0.9$) points toward a peculiar ring geometry as described by Donati et al. (2001).  However, such a ring-model is supposed to be thick in the X-ray band, which implies a strong rotational modulation in the X-ray emission, but it is not observed (Favata et al. 2009). Therefore, if the model of Donati et al. (2001) is valid, the ring should be optically thin even in the X-ray band. In addition, the best-fit geometry we obtain for the ring is somewhat greater than the values provided in Donati et al. (2001): $74 \pm 7 R_{\star}$ (compared to $2R_{\star}$) for the inner ring diameter, and $5 \pm 6 R_{\star}$ (compared to $6R_{\star}$) for the width. 

However, the angular diameter estimate of $\beta$ Cep we obtain ($\Phi_\star=\Phi_{\mathrm{UD}}[V]=0.22\pm0.05$ mas) - even if it is indeed model-dependent - is quite precise (23\% of relative precision). Considering $T_{\mathrm{eff}}=26000$K and $\log g=4$ for $\beta$ Cep (Gies \& Lambert 1992), we derive the linear limb-darkening coefficient of D\'iaz-Cordov\'es et al. (1995) and find $\Phi_{\mathrm{LD}}[V]=0.23\pm0.05$ mas . This value is consistent with the value ($\Phi_{\mathrm{LD}}[V]=0.29\pm0.06$ mas) predicted by Donati et al. (2001) when assuming a distance of $d = 210 \pm 13 pc$ (Van Leeuwen 2007).  In addition, the {\it LITpro} software provides the correlation matrix for all parameters.  It shows that the angular diameter is poorly correlated to other parameters ($<0.5$), which is a good point. This observational value should be compared to the one provided by classical asteroseismological study, using a method similar to the one performed on the solar-like star $\alpha$ Cen A and B (Th\'evenin, F. et al. 2002, Kervella et al. 2003). 

The hypothesis of a spherical shell of X-ray emitting plasma around the star located between 5$R_{\star}$  and 7$R_{\star}$ (Favata et al. 2009) is not excluded. Our circular ring model provides a $1.4$ reduced $\chi^2$ with only 5 parameters; however, the star is again unresolved, which is not expected (Donati et al. 2001). If we consider the angular diameter derived from the peculiar disk-fitted model ($0.22\pm0.05$mas), the values in stellar radii we obtain for the circular ring geometry are more than in Favata et al. (2009): $24 \pm 17 R_{\star}$ for the inner ring diameter and $12 \pm 1 R_{\star}$ for the width. 

Additional observations of $\beta$ Cep in the next years with the VEGA/CHARA instrument will help complete the (u,v) plane coverage and lead to a high-precision determination of the geometrical structure around this star. 
Moreover, high spectral resolution observations in the metallic lines will help characterize the dynamical structure of the atmosphere of $\beta$ Cep (Nardetto et al. 2005), and bring constraints on the stellar inclination axis, a parameter that is fundamental to mode identification (Telting et al. 1997). 

There are 15 bright $\beta$ Cephei stars ($V < 7$)  with expected angular diameter $>0.2$ mas that could be observed by long baseline interferometry in the future, and some of them are binaries (Stankov \& Handler 2005). Such a survey could constrain the close environment of pulsating stars and help for understanding the close interaction between the pulsation, the mass, and the mass loss (and eventually the magnetic field). This would help the future development of the hydrodynamical codes of pulsating stars (e.g. Fokin et al. 2004), as well as of stellar interior models.

\begin{acknowledgements}
VEGA is a collaboration between CHARA and the Laboratoire
Fizeau (OCA/UNS/CNRS-Nice), LAOG in Grenoble, CRAL in Lyon, and
LESIA in Paris-Meudon. It has been supported by French programs for stellar
physics and high angular resolution PNPS and ASHRA, by INSU-CNRS, and by
the R\'egion PACA. The project has benefited from the strong support of the OCA
and CHARA technical teams. The CHARA Array is operated with support from the National Science
Foundation through grant AST-0908253, the W. M. Keck Foundation, the
NASA Exoplanet Science Institute, and from Georgia State University.This research has made use of the Jean-Marie Mariotti Center \texttt{LITpro}\footnote{LITpro software available at http://www.jmmc.fr/litpro} and \texttt{SearchCal} services\footnote{Available at http://www.jmmc.fr/searchcal} services co-developed by CRAL, LAOG, and FIZEAU, and of CDS Astronomical Databases SIMBAD and VIZIER. 

\end{acknowledgements}


\end{document}